%
%
%
%
%
\font\tenmib=cmmib10 \font\sevenmib=cmmib10 scaled 700
\font\fivemib=cmmib5 scaled 500
\skewchar\tenmib'177 \skewchar\sevenmib'177 \skewchar\fivemib'177
\newfam\mibfam  
\textfont\mibfam=\tenmib \scriptfont\mibfam=\sevenmib
\scriptscriptfont\mibfam=\fivemib
%
\def\boldtheta{\mathchar"0812 }
\def\boldpsi  {\mathchar"0820 }

\def\boldomega{\mathchar"0821 }
%
%
\def\bold{\bf}

\def\roman{\rm}
\def\gap {\vskip 6pt}
\def\indenton {\parindent=20pt}
\def\indentoff {\parindent=0pt}
\def\ref#1#2#3#4#5#6{[#1] #2, {\sl #3}, {\bf #4}, #5, (#6).\vskip 6pt
\par}
\magnification=\magstep1
\vsize=25 true cm
\goodbreak
\indenton
\vskip 1cm
\centerline {\bf MAGNETIC DIPOLE ABSORPTION OF RADIATION}
\centerline {\bf IN SMALL CONDUCTING PARTICLES}
\vskip 1 cm
\centerline {Michael Wilkinson\footnote{${^1}$}
{Permanent address: Department of Physics and Applied Physics,
John Anderson Building, University of Strathclyde, Glasgow,
G4 0NG, Scotland, U.K.}
and Bernhard Mehlig\footnote{${^2}$}
{Permanent address: Max-Planck-Institut f\"ur Physik komplexer
Systeme, N\" othnitzer Str. 38, 01187 Dresden, Germany}}
\gap\gap
\centerline {Isaac Newton Institute for Mathematical Sciences,}
\centerline {University of Cambridge,}
\centerline {20, Clarkson Road,}
\centerline {Cambridge, CB3 0EH,}
\centerline {United Kingdom,}
\vskip 1 cm
\centerline {Paul N. Walker}
\gap\gap
\centerline {Laboratoire de Physique des Solides, associ\' e au CNRS}
\centerline {Universit\' e Paris-Sud}
\centerline {91405 Orsay}
\centerline {France}
\vskip 2cm
\centerline{\bf Abstract}
\gap
We give a theoretical treatment of magnetic dipole absorption
of electromagnetic radiation in small conducting particles,
at photon energies which are large compared to the single particle
level spacing, and small compared to the plasma frequency. 
We discuss both diffusive and ballistic electron dynamics
for particles of arbitrary shape.

The conductivity becomes non-local when the frequency is smaller
than the frequency $\omega_{\rm c}$ characterising the transit 
of electrons from one side of the particle to the other, but 
in the diffusive case $\omega_{\rm c}$ plays no role in determining
the absorption coefficient. In the ballistic case, the absorption
coefficient is proportional to $\omega^2$ for $\omega \ll \omega_c$,
but is a decreasing function of $\omega $ for 
$\omega\gg \omega_{\rm c}$.
\gap\gap\gap
\vfill
\eject
%
%
%
%
\noindent{\bf 1. Introduction}
\gap
For sufficiently low frequencies, the interaction of small conducting
particles with electromagnetic radiation is dominated by absorption
rather than scattering. The classical theory of the interaction of
electromagnetic radiation with spherical particles of uniform composition
was considered by Mie [1]; the theory encompasses conducting
particles with a complex dielectric constant, which is often modelled
by the Drude theory [2]. If the particles are small compared to both the
wavelength and the electromagnetic skin depth of the radiation,
the dominant contributions to the absorption are called the
electric and magnetic dipole terms [3].
In both cases the absorption is due to Joule or Ohmic heating
caused by electrical currents flowing through the particle:
the electric dipole term is due to currents which establish
electrical polarisation of the particle, and the magnetic dipole
term is due to eddy currents induced by variation of the magnetic
field.

At frequencies below the plasma frequency, the electric field is
screened from the interior of the particle, but the magnetic field
can penetrate the whole of the particle. Although magnetic effects
are negligible in atomic absorption processes, they could become
significant when the number of atoms in the particle sufficiently large
that most of the atoms are screened from the electric field. In fact, 
magnetic dipole absorption is often the dominant absorption process
in suspensions of small metal particles [4]. Very few of the
many theoretical papers on absorption of radiation by small particles
have considered the magnetic dipole contribution, some exceptions are
[5,6] which consider magnetic dipole absorption in the context
of effective medium theories. Because it is typically the dominant
contribution, it is appropriate to consider the problem of
magnetic dipole absorption in some detail. 

The Mie theory is restricted to spherical particles in which the
dynamics of the charge carriers is diffusive, and quantum mechanical
effects are ignored. In this paper we will give the first theoretical
treatment of magnetic dipole absorption going beyond the Mie theory:
we will consider both diffusive electron dynamics (for which the
particles are characterised by their bulk conductivity), and
ballistic electron motion (in which the particle is smaller than the
bulk mean free path of the conduction electrons). Our approach
allows for arbitrary particle geometries, and we give careful 
consideration to the fact that the conductivity is non-local
when the particles are very small: quantum mechanical
effects are included using a semiclassical approach. The paper
complements [7-9], which gave a comparably comprehensive treatment
of electric dipole absorption. 

We do not explicitly consider the 
structures in the absorption close to the single particle level spacing
which were originally considered by Gorkov and Eliashberg [10].
These structures are determined by repulsion between energy levels
and the appropriate tool to analyse them is random matrix theory.
Their full characterisation requires an estimate of a 
mean-square matrix element, which was not given correctly in [10].
Our results for the low-frequency limit provide the correct estimate 
of this quantity for magnetic dipole absorption.

Sections 2 to 4 will be concerned with various aspects of the
formulation of the problem, discussing respectively the
relation between the absorption of radiation and correlation
functions of the electron motion, the criteria for a self-consistent
solution of the equations determining the electric field driving 
the eddy currents, and the definition and semiclassical estimation 
of the non-local conductivity which is required to determine the 
self-consistent field. Our semiclassical estimates for the non-local
conductivity are closely related to expressions given by Argaman [11].
In section 5 we develop the theory for magnetic 
dipole absorption in particles with diffusive electron motion,
assuming that the electric field is known. This calculation
also yields the form of the non-local conductivity applicable to the
diffusive case: in section 6 we consider the solution for the 
self-consistent field, and discuss results for some specific 
geometries. Our formula for the non-local conductivity is
identical to one given by Serota and co-workers [12,13],
who used diagrammatic techniques. Our derivation is more direct
and requires fewer assumptions: we discuss this point further
in section 6.

Our approach uses a semiclassical estimate described in [14], which 
relates mean squared matrix elements to classical correlation 
functions. We might expect that there should be features in
the absorption spectrum which are related to the characteristic
timescale for decay of classical correlations, in this case 
the typical time for a particle to cross the particle: the 
importance of this timescale was emphasised by Thouless [15], 
and in the case of diffusive electron motion, we will refer to 
the characteristic frequency scale $\omega_{\rm c}=D/a^2$ as the 
Thouless frequency ($D$ is the diffusion constant and $a$ is 
the characteristic size of the particle). Another reason for 
expecting $\omega_{\rm c}$ to play a role in determining the absorption 
coefficient is that the conductivity is non-local when $\omega $ 
is not large compared to $\omega_{\rm c}$. We find however that 
$\omega_{\rm c}$ plays no role in the final expression for the absorption
coefficient. 

We consider the case of ballistic electron motion
in section 7. We are only able to gain
limited information about this case: we find that
the absorption coefficient is proportional to
$\omega^2$ at frequencies small compared to
$\omega_{\rm c}=v_{\rm F}/a$, and that it is a decreasing
function of frequency for $\omega \gg \omega_{\rm c}$.

Our conclusions are in many ways parallel to those
for electric dipole absorption. In ballistic systems, it
was found [7,8] that the electric dipole absorption has 
resonances in the absorption coefficient with a frequency 
scale $\omega_{\rm c}=a/v_{\rm F}$, where $v_{\rm F}$ is the velocity at the 
Fermi energy. By contrast, in the case of diffusive electron 
motion, it was found [9] that there is no structure in the absorption 
coefficient at the frequency scale $\omega_{\rm c}=D/a^2$. 

Finally, we remark that there 
is a large literature concerned with the effects of time
dependent magnetic fluxes on metallic loops: when the
magnetic flux varies sinusoidally, the absorption of energy
by the loop is a special case of the magnetic dipole
absorption which we consider here. Most of the papers on this
topic are concerned with quantum size effects analogous to
those considered by Gorkov and Eliashberg [10]: two recent 
works in this area are [16] and [17].
\gap\gap\gap
%
%
%
%
\vfill
\eject
\noindent{\bf 2. Formulation of the problem}
\gap
The absorption of radiation is usually described by an extinction
coefficient $\gamma (\omega)$, which is defined as the fractional
loss of intensity per unit length of sample, divided by the volume
fraction $F$ occupied by the particles. We will express our results
in terms of the rate of absorption of energy $\langle dE/dt\rangle$
within a single particle. If the amplitude of the electric and magnetic 
fields are ${E}_0$ are $B_0$ respectively, the intensity of the radiation 
is $I={1\over 2}\epsilon_0\,{E}_0^2={1\over 2}\mu_0B_0^2$, and the 
relationship between $\gamma $ and $\langle dE/dt\rangle$ is therefore
$$\gamma ={2\over{V\epsilon_0 c\,{E}_0^2}}
\biggl\langle {dE\over {dt}}\biggr\rangle,\eqno(2.1)$$
where $V$ is the volume of a single particle.
In this paper we will define the absorption coefficient
$\alpha (\omega )$ as the rate of absorption of energy
for a single particle, divided by the electric field intensity:
$$\alpha (\omega)=
{1\over{{E}_0^2}}
\bigg\langle{dE\over{dt}}\bigg\rangle
={1\over{c^2{B}_0^2}}
\bigg\langle{dE\over{dt}}\bigg\rangle
\ .\eqno(2.2)$$
The normalisation with respect to electric (rather than magnetic) 
field intensity is used to facilitate comparison with the results
in [7-9].   

The particle will be considered to consist of a static potential
well which traps a gas of non-interacting fermions (electrons), initially
with occupation probability $f(E)$ (which would be identified with
the Fermi-Dirac distribution). In a quantum mechanical calculation the 
rate of absorption of energy is determined by the Fermi golden rule.
This states that the rate of transition under the action of a
periodic perturbation with frequency $\omega $ and matrix elements
$\Delta H_{nm}$, from a state with energy $E_n$, to a quasi-continuum of
final states with energies close to $E_m=E_n+\hbar \omega$ is
$$R={\pi\over{2\hbar}}g(E_m)\langle \vert 
     \Delta H_{nm}\vert ^2\rangle_\omega
\eqno(2.3)$$
where $g(E_m)$ is the density of final states and 
$\langle \vert \Delta H_{nm}\vert ^2\rangle_\omega$ is the mean-square matrix
element for transitions from $E_n$ to states close to
$E_m$. The energy absorbed by an electron making an upward
transition is $\hbar \omega$. The rate of absorption of energy
is therefore
$$\biggl \langle {dE\over {dt}}\biggr \rangle
=\hbar \omega \int \!dE\, g(E) \,R(E) \,[f(E)-f(E-\hbar \omega)]\sim
Rg \hbar^2\omega^2\eqno(2.4)$$
where the approximate equality is applicable in the limit
where $\hbar \omega$ and $kT$ are both large compared to the mean level 
spacing, but small compared to other energy scales: in the right hand 
expression both $g $ and $R$ are evaluated at the Fermi energy. 

The mean-square matrix element can be estimated semiclassically [14]:
$$\langle |\Delta H_{nm}|^2\rangle_\omega ={1\over {2\pi \hbar g}}
\int_{-\infty}^\infty dt\ \exp(i\omega t)
\langle \Delta H(t)\Delta H(0)\rangle_E\eqno(2.5)$$
where $\langle \Delta H(t)\Delta H(0)\rangle_E$ 
is the microcanonical autocorrelation
function of the classical observable corresponding to 
$\Delta \hat H$, evaluated at energy $E$. Combining this result with 
(2.4), the absorption coefficient can be expressed in terms of the
classical autocorrelation function
of the perturbation $\Delta H({\bold r},{\bold p})$. The resulting expression
can also be written in terms of the classical change in energy of the
individual electrons due to the perturbation: the change in energy
of an electron following a trajectory ${\bold r}(t),{\bold p}(t)$ is
$$\Delta E(t)=\int_0^t dt'\ 
{\partial H\over {\partial t}}({\bold r}(t'),{\bold p}(t'))
\ .\eqno(2.6)$$
Combining this result with (2.5), the rate of absorption of energy
by the electron gas is
$$\biggl \langle {dE\over {dt}}\biggr \rangle={\textstyle g\over 2}\,
{d\over dt}\Big\langle \Delta E(t)^2\Big\rangle
\eqno(2.7)$$
where $\langle \Delta E(t)^2\rangle$ is the variance of the change
in the single-electron energies.

In our problem the perturbation is a sinusoidally varying
electromagnetic field, specified by a vector potential
${\bold A}({\bold r})\exp(i\omega t)$, and a scalar potential
$\Phi ({\bold r})\exp(i\omega t)$. The component of the perturbation
of the Hamiltonian which is quadratic in ${\bold A}$ can be neglected
when calculating the leading order absorption coefficient; the
remaining terms are
$$\Delta \hat H={e\over {2m_e}}
({\bold \hat p}.{\bold A}+{\bold A}.{\bold \hat p})+e\Phi 
\ .\eqno(2.8)$$
In discussing the magnetic dipole absorption, it is given that
the magnetic field is
$${\bold B}(t)=\nabla \wedge {\bold A}={\bold B}_0 \exp[{\rm i}\omega t]
\ .\eqno(2.9)$$
The fluctuating magnetic field induces an electric field ${\bold E}$, 
which is given by
$${\bold E}=-{\partial {\bold A}\over{\partial t}}+\nabla \Phi
\ .\eqno(2.10)$$
The Hamiltonian admits a set of gauge transformations
$({\bf A},\Phi)\to ({\bf A}',\Phi')=({\bf A},\Phi)+
(\nabla \mu,\partial_t\mu)$ which leave the electric
and magnetic fields unchanged. We will assume that the gauge
has been chosen so that $\Phi =0$.
Physically, the electric field is not uniquely defined by the
magnetic field, and must be determined by a self-consistent
condition, which we discuss in section 3.
\gap\gap\gap
\vfill
\eject
%
%
%
%
\noindent{\bf 3. Self-consistent choice of the vector potential}
\gap
\noindent{\sl 3.1 Self-consistent electric field}
\gap
The eddy currents induced by the fluctuating magnetic
will themselves generate a magnetic field. We will consider
only the case of extremely small particles, for which
this additional magnetic field is negligible. We therefore
assume that the magnetic field is simply the externally
applied field, ${\bold B}(t)=B_0\exp [i\omega t]{\bold e}_3$.
The electric field is required to uniquely determine the
perturbation of the Hamiltonian. In this section we will
discuss the self-consistent calculation of the electric
field.

In discussions of the Zeeman effect in atomic physics, the
perturbation of the Hamiltonian representing the magnetic
field is conventionally taken to be proportional to the component
of the angular momentum operator along the direction of the field.
It is natural to ask why a more involved procedure is used
here, but is unnecessary for the Zeeman problem or for
calculation of static magnetic susceptibility. At the end
of this section we show that the angular momentum operator
gives the correct answer in the limit where the frequency
approaches zero, but not in general.

The electric field
satisfies the Maxwell equations
$$\nabla\wedge {\bold E}={\partial {\bold B}\over {\partial t}},\ \ \ \
\nabla .\, {\bold E}={\rho \over {\epsilon_0}}
\ .\eqno(3.1)$$
The electric field causes a current density ${\bold j}$ to flow
within the particle. We will assume that a linear response
theory is valid, but in general the current may be a non-local
function of the electric field: we will write
$${\bold j}({\bold r},t)=\int d{\bold r}'\int_{-\infty}^t \!\!\!dt'\,
\tilde \sigma ({\bold r},{\bold r}';t-t')\,{\bold E}({\bold r}',t')
\eqno(3.2)$$
where $\tilde \sigma$ is the non-local conductance tensor.
We will also write this relation in the form
$${\bold j}({\bold r},\omega)=\int d{\bold r}'\,
\tilde \sigma ({\bold r},{\bold r}';\omega)\,
{\bold E}({\bold r}',\omega)
\equiv\hat \sigma \, {\bf E}({\bf r},\omega)
\eqno(3.3)$$
where the second equality defines an operator $\hat \sigma$ which
maps the electric field ${\bf E}({\bf r},\omega)$ non-locally
into the current field ${\bold j}({\bf r},\omega)$. 
For a monochromatic perturbation, the charge density is
$$\rho =-{{\rm i}\over \omega}\nabla.\,{\bold j}
\ .\eqno(3.4)$$
Combining these results, we find the following equation
for the electric field
$$\nabla.\,{\bold j}'=0,\ \ \ {\bold j}'
=(\hat \sigma -{\rm i}\omega \epsilon_0){\bold E}
\ .\eqno(3.5)$$
This equation must be supplemented by a boundary condition
in order to uniquely determine the electric field. This
is
$$\hat {\bold n}.\,{\bold j}=0\eqno(3.6)$$
where $\hat {\bold n}$ is a unit vector normal to the boundary: 
this condition represents the fact that charges cannot enter or
leave the sample.
\gap
\noindent {\sl 3.2 Representation of the field in terms of potentials}
\gap
It will be convenient to write the electric field in terms
of a vector and a scalar potential:
$${\bold E}=\nabla \wedge {\boldpsi}+\nabla \phi
\ .\eqno(3.7)$$
We will only consider in detail cases where the field
${\boldpsi}({\bf r})$ is of the form
$${\boldpsi}=\psi (x,y){\bold e}_3
\ .\eqno(3.8)$$
This form is appropriate when the conducting particle is
two dimensional, lying in the plane $z=0$, for three
dimensional particles in the form of general cylinders
aligned with the $z$ axis, and can be extended to
spheres and some other non-cylindrical geometries. 
Substituting (3.7), (3.8) into the Maxwell equations, we 
find that $\psi$ satisfies Poissons's equation in the form
$$\nabla^2 \psi={\rm i} \omega B_0
\ .\eqno(3.9)$$
We will always choose $\psi(x,y)$ to satisfy the condition 
that $\psi=0$ on the boundary. Having uniquely specified
$\psi (x,y)$, equations (3.5) and (3.6) are transformed 
into a equations determining the scalar potential $\phi $.

In some cases a local, isotropic conductivity $\Sigma(\omega)$
will provide an adequate description. In this case, the condition
(3.5) reduces to the requirement that $\nabla. {\bold E}=0$, in
which case the electric field can be written in the form
$${\bold E}=\nabla \wedge {\boldpsi }+\nabla \phi, \ \ \  
\nabla^2 \phi =0
\ .\eqno(3.10)$$
The boundary condition corresponding to (3.6) is
then satisfied by taking a solution for which $\psi =0$ and 
$\phi =0$ on the boundary: the latter condition implies 
that $\phi =0$ everywhere.
\gap
\noindent{\sl 3.3 A remark on the low-frequency limit}
\gap
After having described the approach used to define the correct
perturbation, we will now show that any form of the electric field
which has a uniform value of $\nabla \wedge {\bold E}$ gives
the correct value of the absorption coefficient in the limit 
$\omega \to 0$. According to (2.7), the absorption coefficient
is proportional to the variance of the change in the single
particle energy. The change of the single-particle energy
can be written
$$\Delta E(t)=\int \!d{\bold r}.\,{\bold E}
=\int_{t_0}^t \!dt'\ {d{\bold r}\over{dt'}}.\,{\bold E}({\bold r}(t'))
={\rm i}\omega\int_{t_0}^t \!dt'\  {d{\bold r}\over {dt'}}.\,
{\bold A}({\bold r}(t'))
\ .\eqno(3.11)$$
Consider the effect of making a transformation
${\bold A}\to {\bold A}'={\bold A}+\nabla \varphi$ on the
absorption coefficient. The change in the single particle
energy is transformed to $\Delta E'$:
$$\Delta E'=\Delta E+{\rm i}\omega
\int^t dt'\, \exp ({\rm i}\omega t'){d{\bold r}\over {dt}}.
\nabla \varphi $$
$$=\Delta E+{\rm i}\omega\int_{t_0}^t d\varphi [{\bold r}(t')]\, 
\exp({\rm i}\omega t')\equiv \Delta E+{\rm i}\omega X(t,\omega)
\ .\eqno(3.12)$$
If $\omega =0$, the correction $X(t)$ introduced by the 
transformation is simply
$$X(t,0)=\int_{t_0}^t dt'\ {d{\bold r}\over{dt}}.\nabla \varphi
=\varphi[{\bold r}(t)]-\varphi[{\bold r}(t_0)]\eqno(3.13)$$
which remains bounded as $t\to \infty$. For finite
$\omega $ the correction $X(t,\omega)$ satisfies
$$\langle X^2(t,\omega)\rangle=
t\int_{-\infty}^\infty d\tau \ \exp({\rm i}\omega t)
\biggl\langle {d\varphi\over{dt}}(\tau){d\varphi\over{dt}}(0)\biggr\rangle
+O(1)\equiv \Gamma(\omega)t+O(1)
\eqno(3.14)$$
provided the correlation function of $d\varphi /dt$ decays faster
than $Ý\tauÝ^{-1}$. Comparison with (3.13) shows that the coefficient 
$\Gamma(\omega)$ approaches zero as $\omega \to 0$, implying that the 
gauge dependent contribution to the absorption coefficient vanishes 
in the limit $\omega \to 0$.
\gap\gap\gap
\vfill
\eject
%
%
%
%
\noindent{\bf 4. Semiclassical theory for non-local conductivity}
\gap
\noindent{\sl 4.1 General formula}
\gap
We will use a semiclassical analysis for the non-local
conductivity $\tilde \sigma ({\bold r},{\bold r}',\omega)$.
We will first consider the problem in rather abstract
terms: we will discuss a Hamiltonian $H({\bold r},{\bold p},X)$,
where $X$ is a time-dependent parameter. The Hamiltonian
determines the motion of particles in a gas with phase space
density $\rho ({\bold r},{\bold p},t)$. The phase space density
satisfies the Liouville equation $\partial_t \rho=\{\rho ,H\}$.
A solution can be written in the form
$$\rho({\bold r},{\bold p};t)=f (H({\bold r},{\bold p};X)-E_{\rm F})$$
$$-\dot X\int_{-\infty}^t dt'\ {\partial H\over {\partial X}}
({\bold r}(t'),{\bold p}(t');X)\, {\partial f\over{\partial E}}
(H({\bold r},{\bold p};X-E_{\rm F})+O(\dot X^2)
\ .\eqno(4.1)$$
Formally, this is an expansion in the velocity
of the perturbation, $\dot X$: the results will be 
valid for all frequncies, because the amplitude of 
the perturbation is infinitesimal.

The leading order Weyl or Thomas-Fermi estimate of the density 
of states of a quantum system states that the density of quantum 
states is $(2\pi \hbar)^{-d}$ in classically accessible regions of
phase space. We will therefore multiply the above solution by
this factor, and take $f(E)$ to be the Fermi-Dirac function.

Our Hamiltonian, $H=({\bold p}-e{\bold A})^2/2m_{\rm e}+V$ has a 
time dependent vector potential, so that we can write
$$\dot X{\partial H \over{\partial X}}
={\partial H\over \partial{{\bold A}}}.\,{\bold E}
={e\over m_{\rm e}}{\bold p}.\,{\bold E}
\ .\eqno(4.2)$$
The resulting current is
$${\bold j}({\bold r},t)={e\over m_{\rm e}}\int d{\bold p}\ {\bold p}\,
\rho ({\bold r},{\bold p},t)
\ .\eqno(4.3)$$
The current flowing in response to the electric field
${\bold E}({\bold r},t)$ is therefore
$$j_i({\bold r},t)={e^2\over{(2\pi \hbar)^d m_{\rm e}^2}}\sum_j
\int_{-\infty}^t dt'\ \int d{\bold p}\ p_i\ 
{\partial f\over{\partial E}}\ 
P_j({\bold r},{\bold p};t'-t)\ 
E_j({\bold R}({\bold r},{\bold p};t'-t)$$
$$={e^2\over{(2\pi \hbar)^dm_{\rm e}^2}}\sum_j \int_{-\infty}^t \!\!\!\!dt'
\int d{\bold p}\int d{\bold r'}\ {\partial f\over{\partial E}}\ 
p_i\ P_j({\bold r},{\bold p};t'-t)\ 
\delta[{\bold r}'-{\bold R}({\bold r},{\bold p};t-t')]\ 
E_j({\bold r}',t')
\eqno(4.4)$$
where $P_i({\bold r},{\bold p};\tau)$, $R_i({\bf r},{\bf p},\tau)$ are 
the $i^{\roman{th}}$ components of the momentum and position at 
time $\tau $ for a trajectory which starts
at $({\bold r},{\bold p})$ at time $t=0$. The components of the 
non-local conductivity tensor are therefore
$$\sigma_{ij}({\bold r},{\bold r}';t)=
{e^2\over{(2\pi \hbar)^d m_{\rm e}^2}}\int d{\bold p}\int d{\bold r'}
{\partial f\over{\partial E}}\ p_i\ P_j({\bold r},{\bold p};t')
\,\delta[{\bold r}'-{\bold R}({\bold r},{\bold p};t-t')]\ .\eqno(4.5)$$
We can write this result in a simpler form
$$\sigma_{ij}({\bf r},{\bf r}',t)={e^2\over{(2\pi \hbar)^d m_{\rm e}^2}}
\bigl\langle p_i({\bf r},t)p_j({\bf r}',0)\bigr\rangle
\eqno(4.6)$$
where the angle brackets denote an average over the initial 
momenta, defined by (4.5). 
We will consider the evaluation of this quantity for diffusive motion
in section 6; next we consider the case of ballistic motion.
\gap
\noindent{\sl 4.2 Results specific to ballistic systems}
\gap
Equation (4.5) can also be expressed as a sum over classical
trajectories which travel between ${\bold r}$ and ${\bold r'}$:
$$\sigma_{ij}({\bf r},{\bf r}',t)=
{e^2\over{(2\pi \hbar)^dm_{\rm e}^2}}\sum_{\roman{paths}}
\Bigg[{\roman{det}}\biggl({\partial R_k\over{\partial p_l}}\biggr)\Bigg]^{-1}
(p_{\roman{init}})_j(p_{\roman{fin}})_i\ .\eqno(4.7)$$
In the low temperature limit the term $\partial f/\partial E$
reduces to a delta function, and this expression becomes a sum
of delta functions $\delta (t-\tau_j)$, where the $\tau_j$ are
the times of trajectories from ${\bold r}$ to ${\bold r}'$ at the
Fermi energy $E_{\rm F}$.

It is more convenient to consider the frequency dependent
non-local conductivity: if the electric field is
${\bold E}({\bold r})\exp({\rm i}\omega t)$, then the current
can be written in the form
$$j_i({\bold r},t)=\exp({\rm i}\omega t)\sum_j \int d{\bold r}'
\sigma_{ij}({\bold r},{\bold r}';\omega )\ E_j({\bf r}')\eqno(4.8)$$
Comparing with (4.4) and (4.5), we find
$$\sigma_{ij}({\bold r},{\bold r}';\omega)=\int d\tau\ 
\exp({\rm i}\omega \tau)\ \sigma_{ij}({\bold r},{\bold r}';\tau)$$
$$={e^2\over{(2\pi \hbar)^2 m_{\rm e}^2}}\sum_{\roman{paths}} \int d\tau \
\exp({\rm i}\omega \tau)\int d\theta \ p_i(\theta)\ p_j'(\theta)\ 
\delta({\bold r}'-{\bold R}({\bold r},{\bold p};\tau))\eqno(4.9)$$
where in the second line we have specialised to the case
of two dimensions, and $\theta $ is the initial angle
of the trajectory. Performing the integrations, we find
$$\sigma_{ij}({\bold r},{\bold r}';\omega)=
{e^2\over{(2\pi \hbar)^2 m_{\rm e}^2}}\sum_{\roman{paths}}\biggl[{\roman{det}}
\biggl({\partial^2{\bold R}\over{\partial \tau \partial \theta}}\biggr)
\biggr]^{-1} \,p_i(\theta)\,p_j'(\theta)\,\exp({\rm i}\omega \tau_k)
\eqno(4.10)$$
where the sum runs over all trajectories which travel from 
$\bold r$ to $\bold r'$ in time $\tau_k$ at the Fermi energy.
This general expression can be specialised in a variety of ways.
We remark that it has a rather simple form for billiards with
boundaries consisting of only straight edges. In this case
the times $\tau_k$ are proportional to the lengths $L_k$
of the trajectories, and because there is no focusing or
de-focusing of bundles of trajectories when they bounce
off the boundary, the form of the determinant is very simple:
we find
$$\sigma_{ij}({\bold r},{\bold r}';\omega)={e^2\over{(2\pi \hbar)^2 
m_{\rm e}^2}}
p_{\rm F}
\sum_k L_k^{-1}\ n_i^{(k)}n_{j'}^{(k)}\exp({\rm i}\omega L_k m_{\rm e}/
p_{\rm F})
\eqno(4.11)$$
where $n_i$, $n_j'$ are the components of a unit vector in
respectively the initial and final directions of the trajectory.
As an example, the case of a square billiard is illustrated in
figure 1.  The formula also gives the non-local conductivity in free space,
with only the direct trajectory included.

We close this section by remarking that in the limit $\omega \to \infty$
the non-local conductivity becomes a very rapidly varying function of
${\bold r}'$, except for when the path length of the trajectory is
very short. Unless the electric field is a rapidly varying function
of position, the dominant contribution to (4.4) comes from the
region where ${\bold r}'$ is close to ${\bold r}$, implying that
a local conductivity $\Sigma_{ij}({\bold r},\omega)$ will give
an adequate description. When ${\bold r}'$ is close to ${\bold r}$,
the non-local conductivity can be approximated by (4.11), with
only the direct trajectory included. The local conductivity
is then obtained as follows:
$$\Sigma_{ij}(\omega)={e^2p_{\rm F}\over{(2\pi \hbar)^2m_{\rm e}^2}}
\int d{\bold R}\ {1\over R}\ \exp({\rm i}\omega m_{\rm e} R/
p_{\rm F})\, n_i\, n'_j$$
$$={{\rm i}e^2p_{\rm F}^2\over{2\pi \hbar^2m_{\rm e}\omega}}\delta_{ij}
={{\rm i}Ne^2\over m_{\rm e}\omega}\delta_{ij}\eqno(4.12)$$
where $N$ is the electron density per unit volume.
This result is precisely the same as the high frequency
limit of the Drude formula for the conductivity.
\gap\gap\gap
\vfill
\eject
%
%
%
%
\noindent{\bf 5. Absorption coefficient for diffusive electrons}
\gap
\noindent{\sl 5.1 Preliminary comments}
\gap
In the present section we calculate the absorption coefficient
$\alpha(\omega)$ for systems with diffusive electron motion,
using (2.6) and (2.7), assuming that the self-consistent
electric field is known.
We show that the absorption coefficient can be written as
a sum of two terms. The first term describes a classical bulk
contribution. The second term introduces boundary contributions 
which could modify the absorption coefficient 
at frequencies below $\omega_{\rm c}=D/a^2$.

We begin by briefly discussing the classical expression for the
absorption coefficient: it is natural to compare the final answer
with this result. The rate of absorption of energy is given by
integrating the rate of Joule heating ${\bf j}.{\bf E}$ over the
volume of the particle. The current density ${\bf j}$ is proportional
to the local electric field:
${\bf j}=\Sigma_0 {\bf E}={\rm i}\omega \Sigma_0 {\bf A}$ where
$\Sigma_0$ is the bulk conductivity of the metal. We therefore have
$$\alpha(\omega)={1\over {2\Sigma_0 E_0^2}}
\int \! d{\bf r}\ |{\bf j}|^2
={\Sigma_0\omega^2\over {2E_0^2}}\int \!d{\bf r}\,{\bf A}^2
={ne^2D\omega^2\over{2c^2B_0^2}}\int d{\bf r}\ {\bf A}^2
\eqno(5.1)$$
where $n$ is the density of states per unit volume, $n=dN/dE=g/V$.
The boundary condition for the electric field
is determined by the fact that the current must be
tangential to the boundary. Unless there is
a constant biasing magnetic field present,
${\bf E}$ is aligned with ${\bf j}$. This implies that
${\bf E}.\hat {\bf n} = 0$ at all points on the boundary (where
$\hat {\bf n}$ is a unit vector normal to the boundary).

Equation (2.5) suggests that the absorption coefficient
might exhibit deviations from classical behaviour at
frequencies small compared to the Thouless frequency
$\omega_{\rm c}=a^2/D$, which
is the inverse of the time taken for an electron to diffuse
across the sample. In the following, we give a semiclassical 
treatment of the absorption coefficient with diffusive
electron motion. There are two distinct but related issues
which must be addressed here. Firstly, for a given
electric field ${\bf E}({\bf r})$, does the absorption
coefficient exhibit any structures at the Thouless energy?
Secondly, is the self consistent solution for the electric
field different above and below the Thouless frequency?
In this section we consider the first of these issues.
In section 6 we will show how one of the results
below can be used to determine the non-local conductance, 
and consider the determination of the self consistent
field in greater detail.
\gap
\noindent{\sl 5.2 Calculation of the energy absorbed}
\gap
Our calculation is based upon (2.6) and (2.7). Because the
instantaneous velocity is not well defined for a diffusive
trajectory, we will divide the
trajectory of the electron into finite segments, in which
the electron travels from ${\bf r}_n$ to ${\bf r}_{n+1}$
with a uniform velocity, in a fixed time increment $\delta t$.
The ${\bf r}_n$ are chosen from an ensemble of random walks
confined within the boundary of the particle. 
The change in the single-electron energy is
$$\Delta E(t)={\rm Re}\biggl[
{\rm i}e\omega\int_0^t dt' \exp({\rm i} \omega t)
{\bf A}({\bf r}).{d{\bf r}\over {dt'}}
\biggr]$$
$$={\rm Re}\biggl[
{\rm i}e\omega\sum_{n=0}^{{\cal N}-1}
\exp({\rm i} \omega t_n)\bar {\bf A}({\bf r}_n).\delta{\bf r}_n
\biggr]\eqno(5.2)$$
where $t={\cal N}\delta t$, $t_n=(n+{1\over 2})\delta t$,
$\delta {\bf r}_n={\bf r}_{n+1}-{\bf r}_n$,
and the quantity $\bar {\bf A}$ is defined by the requirement that
each term in the sum equals the contribution to the integral from
the corresponding link in the random walk. 
For diffusive motion with fixed diffusion constant, we 
must take $\langle \delta r^2\rangle \sim \delta t$,
so that the error in each step must be $O(\delta r^3)$
to achieve a convergent estimate of the integral.
The sum (5.2) can be approximated as follows:
$$\Delta E(t)={\rm Re}\biggl[
{\rm i}e\omega\sum_{n=0}^{{\cal N}-1} \exp({\rm i}\omega t_n)
{\bf A}[{\textstyle{1\over 2}}({\bf r}_n+{\bf r}_{n+1})]
.({\bf r}_{n+1}-{\bf r}_n)
\biggr]+O({\cal N}\delta r^3)\eqno(5.3)$$
and the error term vanishes in the limit $\delta t\to 0$.
We remark that equation (5.2) is a stochastic integral, and evaluation 
of the integrand of the midpoint of the step is equivalent to using
the Stratonovich definition of the integral [18].
The variance of (5.3) is
$$\langle \Delta E(t)^2\rangle
=e^2\omega^2\;\sum_{i,j=0}^{{\cal N}-1}\exp[{\rm i} \omega (t_i-t_j)]
\langle
{\bf A}[{\textstyle{1\over 2}}({\bf r}_i+{\bf r}_{i+1})].\delta {\bf r}_i\,
{\bf A}[{\textstyle{1\over 2}}({\bf r}_j+{\bf r}_{j+1})].\delta {\bf r}_j
\rangle+O(\delta t)$$
$$\sim e^2 \omega^2 {\cal N} \sum_{n=-\infty}^{\infty} 
\exp({\rm i} \omega t_n)C_n
\eqno(5.4)$$
where
$$C_n=\langle{\bf A}[{\textstyle{1\over 2}}({\bf r}_0+{\bf r}_{1})].
\delta {\bf r}_0\,
{\bf A}[{\textstyle{1\over 2}}({\bf r}_n+{\bf r}_{n+1})].
\delta {\bf r}_n\rangle
\eqno(5.5)$$
The absorption coefficient is therefore proportional to the Fourier
transform of a correlation function $C(t)$:
$$\alpha (\omega)={ge^2\omega^2\over{2\delta t B_0^2}}
\sum_{n=-\infty}^\infty \exp({\rm i}\omega t_n)C_n
\equiv {ge^2\omega^2\over{2B_0^2}}
\int_{-\infty}^\infty dt \exp({\rm i}\omega t)C(t)
\eqno(5.6)$$
where $C(n\delta t)\equiv C_n/(\delta t)^2$.
The correlation function can be expressed in terms of the
propagator ${\cal P}({\bf r},{\bf r}';t)$ which gives the probability
density for reaching ${\bf r}'$ from initial position ${\bf r}$ at 
time $t$: for times large compared to $\delta t$ this satisfies
the diffusion equation $[\partial_t-D\nabla^2_{\bf r}]{\cal P}=0$.
Assuming summation over repeated indices, we find 
$$C(t)={1\over {V\delta t^2}}\int d{\bf r} \int d{\bf r}'
\int d\delta {\bf r} \int d\delta {\bf r}'\,
A_i({\bf r}+{\textstyle{1\over 2}}\delta {\bf r})\,
A_j({\bf r}'+{\textstyle{1\over 2}}\delta {\bf r}')\,
\delta r_i \, \delta r'_j$$
$$\times {\cal P}({\bf r}+\delta {\bf r},{\bf r}';t-\delta t)\,
         {\cal P}({\bf r},{\bf r}+\delta {\bf r};\delta t)\,
         {\cal P}({\bf r}',{\bf r}'+\delta {\bf r}';\delta t)\,.
\eqno(5.7)$$
We discuss the cases $t=0$ and $t\neq 0$ seperately.
At $t=0$, we have $\delta {\bf r}=\delta {\bf r}'$, and the correlation
function is easily evaluated, giving a result which is 
of order $O(\delta t)$:
$$C_0=\langle ({\bf A}.\delta {\bf r})^2\rangle
={2D\delta t\over{V}}\int d{\bf r}{\bf A}({\bf r})^2
\ .\eqno(5.8)$$
Here we have used $\langle ({\bf A}.\delta {\bf r})^2\rangle
\sim {1\over 2}\langle {\bf A}^2 \rangle\langle\delta {\bf r}^2\rangle
= D\delta t \langle {\bf A}^2\rangle$. This result implies that
there is a contribution to $C(t)$ of the form $(D/V)\delta (t)$.

The case $t\ne 0$ requires a more delicate treatment.
We can expand (5.7) in both the short time interval
$\delta t$ and in the short step $\delta {\bf r}$. It will
turn out that the leading order contribution is 
of the order of $O(\delta t^2)$,
as opposed to the case $t=0$. 
Because the motion is diffusive,
we have $\delta {\bf r}^2\sim \delta t$, so that terms up to quartic
in $\delta {\bf r}$ must be retained 
$$C(t)\sim {1\over V\delta t}\int d{\bf r} \int d{\bf r}' 
                 \int d\delta {\bf r} \int d \delta {\bf r}'\,
[A_i+{\textstyle{1\over 2}}\partial_{r_k} A_i\,\delta r_k
+{\textstyle{1\over 8}}\partial^2_{r_{k\phantom'}r_{l}}A_i\,
\delta r_{k}\,\delta r_{l}]$$
$$\times [A'_j+{\textstyle{1\over 2}}\partial_{r'_k} A'_j\,\delta r'_k
+{\textstyle{1\over 8}}\partial^2_{r'_kr'_{l}}A'_j
\,\delta r'_k\,\delta r'_{l}]\ \delta r_i \,\delta r'_j$$
$$\times [{\cal P}({\bf r},{\bf r}';t)-\partial_t 
          {\cal P}({\bf r},{\bf r}';t)\,\delta t
+\partial_{r_k}
          {\cal P}({\bf r},{\bf r}';t)\,\delta r_k
+{\textstyle{1\over 2}}\partial^2_{r_kr_{l}}
          {\cal P}({\bf r},{\bf r}';t)
\delta r_k\delta r_{l}]$$
$$\times {\cal P}({\bf r},{\bf r}+\delta {\bf r};\delta t)\, 
         {\cal P}({\bf r}',{\bf r}'+\delta {\bf r}';\delta t)
\ .\eqno(5.9)$$
The terms containing $\partial_t {\cal P}({\bf r},{\bf r}';t)$ can be 
dropped when there are more than two factors of $\delta {\bf r}$. The 
integrals over products of the $\delta {\bf r}$ can now be separated 
out to give 
$$C(t)={1\over V\delta t^2}\int d{\bf r} \int d{\bf r}'\biggl\{
A_i\,A_j\,[{\cal P}-\partial_t {\cal P}\,\delta t]
\langle \delta r_i\rangle \langle \delta r'_j\rangle$$
$$+[{\textstyle{1\over 2}}A'_j\partial_{r_k}A_i\, {\cal P}
+{\textstyle {1\over 2}}A'_j\partial_{r_k}({\cal P} A_i)-
{\textstyle {1\over 2}}A'_j\partial_{r_k}A_i\, {\cal P}]
\langle \delta r_i \delta r_k \rangle \langle \delta r'_j \rangle$$
$$+[{\textstyle{1\over 2}}A_i\partial_{r_k}A'_j\, {\cal P}]
\langle \delta r'_j \delta r'_k \rangle \langle \delta r_i \rangle$$
$$+[{\textstyle{1\over 8}}A_i\partial^2_{r_kr_{l}}A'_j\,{\cal P}
+{\textstyle{1\over 2}}A_i\,A'_j\partial^2_{r_kr_{l}}{\cal P}]
\langle \delta r_i \delta r_k \delta r_{l}\rangle \langle \delta r'_j
\rangle$$
$$+[{\textstyle{1\over 2}}A_i\partial^2_{r_kr_{l}}A'_j\,{\cal P}]
\langle \delta r'_j \delta r'_k \delta r'_{l}\rangle \langle
\delta r_i \rangle$$
$$+[{\textstyle{1\over 4}}\partial_{r_{l}}A_j\partial_{r_k}A_i\,{\cal P}
+{\textstyle{1\over 2}}A_i\partial_{r_{l}}A_j\partial_{r_k}{\cal P}]
\langle \delta r_i \delta r_k \rangle \langle \delta r'_{l}\delta r'_j
\rangle
\biggr\}
\eqno(5.10)$$
where $\langle \delta {\bf r}\rangle 
= \int \!d\delta {\bf r} \; \delta{\bf r}
\; {\cal P}({\bf r,}{\bf r}+\delta{\bf r};\delta t)$.
Now consider the form of these integrals when $\delta t$ is
sufficiently small. The propagator
${\cal P}({\bf r},{\bf r}+\delta {\bf r};\delta t)$
is small unless $\delta {\bf r}$ is small.
When ${\bf r}$ is not close to the boundary, this propagator
can be approximated by a function of the distance travelled, 
${\cal P}_0(\vert \delta {\bf r}\vert,\delta t)$: because the steps are 
assumed to be independent, the variance $\langle \delta {\bf r}^2\rangle$
averaged over this distribution 
can be identified with $2dD\delta t$ (where $d$ is the dimensionality
of space). When ${\bf r}$ is close to the boundary,
a solution satisfying the boundary condition ${\bf n}.\nabla
{\cal P}=0$ is constructed by the method of images. We denote
the image of the source point ${\bf r}$ by ${\bf r}^\ast$. 
The diffusion propagator is then
$${\cal P}({\bf r},{\bf r}+\delta {\bf r}; \delta t)
\sim {\cal P}_0(\vert \delta {\bf r}\vert,\delta t)
    +{\cal P}_0(\vert \delta {\bf r}+{\bf r}-{\bf r}^\ast \vert,\delta t)
\ .\eqno(5.11)$$
Since $\delta t$ is small compared to the time to traverse the
particle, the second term only contributes for points ${\bf r}$
close to the surface, which can thus be considered locally flat. We
then introduce a local coordinate system arranged so that the nearest
boundary point defines the origin. In two dimensions, the surface
tangent is given by the line $x=0$, and the normal by $y=0$. the
point ${\bf r}$ lies at $(x,0)$, and ${\bf r}^\ast=(-x,0)$ 
(figure 2). 

The average $\langle \delta r_i\rangle$ vanishes unless ${\bf r}$
is close to the boundary, in which case the mean displacement is 
inwards, and its projection in the direction perpendicular
to the surface is
$$\langle \delta x \rangle_x=\int_0^\infty dx' (x'-x)[f(x'-x)+f(x'+x)]
\eqno(5.12)$$
where $f(x)$ is the projection of the 
distribution ${\cal P}_0(\vert \delta {\bf r}\vert)$ onto the $x$ 
axis
$$f(x)=\int d{\bf r}\ {\cal P}_0(\vert \delta {\bf r}\vert,\delta t)
\,\delta (x-\vert \delta {\bf r}\vert)\eqno(5.13)$$
which satisfies
$$\int_{-\infty}^\infty \!dx \, x^2 f(x)=2D\delta t 
\ .\eqno(5.14)$$
Equations (5.12) and (5.14) show that the mean inward displacement 
$\langle \delta x\rangle_x$ is of typical magnitude 
$\sqrt{D\delta t}$, in a layer of depth $\sqrt{D\delta t}$
next to the boundary, and negligible elsewhere. The weight $w$ of 
these inward displacements is clearly $\sim D\delta t$. We define
$$w\equiv \lim_{L \to \infty} \int_0^L \!\!dx \langle \delta x \rangle_x
\ .\eqno(5.15)
$$
We evaluate $w$ by sustituting (5.12), then making a change
of variables $X=x'+x$, $X'=x'-x$. The integral is written as
the sum of two integrals, one over the domain $X\le L$, 
$\vert X'\vert \le X$, which vanishes because of a symmetry,
and another integral which involves only $f(X')$ in the limit
$L\to \infty$: we find $w=D\delta t$, which implies that
$$\int d{\bf r} F_i({\bf r}) \langle \delta {r}_i \rangle =
-D\delta t\,\int d s_i F_i
\eqno(5.16)$$
for any vector field ${\bf F}$, where $ds_i$ are the components 
of a vector element of the surface. When evaluating the integrals
over $\langle \delta r_i\delta r_j \rangle$ we can approximate
these terms by $2D\delta t \delta_{ij}$, because the second term
in (5.14) is significant only in a narrow layer of width
$\sqrt{D\delta t}$. The terms containing averages of
$\delta r^3$ make no contribution at order $\delta t^2$.
Retaining only the leading order terms, we find
the following contribution for $t\ne 0$:
$$C(t)={D^2\over V}\int \!d{\bf r} \int \!d{\bf r}' \,
\partial_{r_i^{\phantom'}} A_i ({\bf r})\,
\partial_{r_j'} A_j ({\bf r}')\, 
{\cal P}({\bf r},{\bf r}';t)$$
$$-2{D^2\over V}\int \!ds_i \!\int \!d{\bf r}' A_i ({\bf r})\,
\partial_{r_j^{\prime}} A_j({\bf r}') \,{\cal P}({\bf r},{\bf r}';t)$$
$$+{D^2\over V}\int \!ds_i \!\int \!ds_j' \,A_i({\bf s})\,
A_j ({\bf s}')\, {\cal P}({\bf s},{\bf s}^\prime;t)\,.
\eqno(5.17)$$
After integrating by parts, and adding the delta function
contribution from $t=0$, we find
$$C(t)=D\delta(t)\,\int \!d{\bf r} \,A_i\,A_i-{D^2\over V}
\int \!d{\bf r}\! \int d{\bf r}' \,A_i\, A_j\, 
\partial^2_{r_ir'_j} {\cal P}({\bf r},{\bf r}';t)
\eqno(5.18)$$
Before Fourier transforming this expression to determine
the absorption coefficient, we will introduce a convenient
expression for the propagator:
$${\cal P}({\bf r},{\bf r}';t)=\sum_\alpha 
\chi_\alpha ({\bf r}) \chi_\alpha ({\bf r}')
\exp(-Dk_\alpha^2t)\eqno(5.19)$$
where the $\chi_\alpha ({\bf r})$ are solutions of the Helmholz
equation $[\Delta + k_\alpha^2]\chi_\alpha=0$,
satisfying the Neumann boundary condition
$n_i\partial_{r_i}\chi_\alpha =0$. Using this result,
we find the following expression for the absorption
coefficient
$$
\alpha(\omega )=K\omega^2\biggl[\int d{\bf r}\, {\bf A}^2-
\sum_\alpha {D^2k^2_\alpha\over {(Dk_\alpha^2)^2+\omega^2}}
\bigg\vert\int d{\bf r}\, A_i\partial_{r_i}\chi_\alpha \bigg\vert^2
\biggr]
\eqno(5.20)
$$
where
$$K={ge^2D\over{B_0^2}}={\Sigma_0\over{B_0^2}}
\ .\eqno(5.21)$$
Equation (5.20) is the main result of this section. As pointed out
above, it consists of two terms. The first term is just 
the classical result derived at the beginning of this section.
The second term is written as a sum over eigenmodes of
the diffusion propagator. It incorporates boundary effects,
as will be seen in the next section. 
\gap
\noindent{\sl 5.3 Low-frequency limit}
\gap
We conclude this section with two remarks concerning the
low frequency limit. First, we show that
that the result (5.20) fulfils a condition
discussed at the end of section 3, where we noted that the
low-frequency limit of the absorption coefficient must be
invariant under adding any gradient to the vector potential.
To see this, consider
the effect of the following transformation:
$$A_i\to A_i+\partial_{r_i}\varphi\,.\eqno(5.22)$$
In the limit $\omega\rightarrow 0$ the absorption coefficient
is determined by the integral of the correlation
function. The addition of the term $\partial_{r_i} \varphi $
produces two new terms in the delta function contribution
in (5.18),
one an integral containing $A_i\partial_{r_i}\varphi$, the
other containing $\partial_{r_i}\varphi \partial_{r_i}\varphi $.
Corresponding additional terms appear in the double
space integral in (5.18). Consider the first of these terms,
linear in $\partial_{r_i}\varphi $. In order for 
these terms to make no contribution
to the absorption coefficient, it is sufficient to show that
the following two integrals are equal
$$I=D\int_{-\infty}^\infty \!\!dt\, \int\! d{\bf r}\! 
                                    \int\! d{\bf r}'\,
A_i \,\partial_{r_j}\varphi \,\partial^2_{r_i,r'_j}{\cal P}
\eqno(5.23)$$
$$I'=2\int dr A_i \partial_{r_i}\varphi\,.
\eqno(5.24)$$
We will use the result
$$\delta({\bf r}-{\bf r}')=
\sum_\alpha \chi_\alpha ({\bf r})\chi_\alpha ({\bf r}')\eqno(5.25)$$
which follows from setting $t=0$ in (5.19). Using (5.20), we find
$$I=2\sum_\alpha {1\over {k_\alpha^2}}
\int \!d{\bf r} \int d{\bf r}' 
A_i \partial_{r_j}\varphi\,
\partial_{r_i}\chi_\alpha({\bf r}) \,
\partial_{r_j}\chi_\alpha({\bf r}')
$$
$$=2\int d{\bf r} \int d{\bf r}' 
A_i \varphi ({\bf r}')\,
\partial_{r_i}\chi_\alpha({\bf r}) \,
\chi_\alpha ({\bf r}')$$
$$=2\int d{\bf r} \int d{\bf r}' 
A_i \,\varphi ({\bf r}') \,
\partial_{r_i'}\delta ({\bf r-r'})=I'\,.
\eqno(5.26)$$
This shows that terms involving $A_i\partial_{r_i} \varphi$ cancel
and do not contribute to the absorption in the limit $\omega \to 0$.
Since this result applies for any vector field $A_i$, we
can replace $A_i$ by $\partial_{r_i}\varphi$ and deduce
immediately that the terms quadratic in $\partial_{r_i}\varphi $ 
also cancel.

Second, we comment on the form of the absorption coefficient
in the limits $\omega \gg \omega_{\rm c}$ and $\omega \ll \omega_{\rm c}$.
The vector potential can always be written as a sum of
the curl of a divergenceless field, and a gradient
$$
{\rm i}\omega{\bf A}
={\bf a}+\nabla \varphi=\nabla \wedge {\boldpsi}+\nabla \varphi 
\ ,\ \ \
\nabla .\,{\boldpsi}=0
\eqno(5.27)$$
with the field ${\bf a}$ chosen so that it is tangential
to the boundary ($\hat {\bf n}.{\bf a}=0$). 
In the case $\omega \gg \omega_{\rm c}$,
the absorption coefficient is determined by
the delta function contribution, and we have
$$\lim_{\omega /\omega_{\rm c} \to \infty}\alpha(\omega)
=K\omega^2 \int dr\ \big( {\bf a}^2 +\nabla \varphi^2\big)
\eqno(5.28)$$
with $K$ given by (5.21).
In the case $\omega \ll \omega_{\rm c}$, on the other hand,
the calculation we described
above shows that the potential $\varphi $ makes no contribution
to the absorption coefficient, and that
$$\lim_{\omega /\omega_{\rm c} \to 0}\alpha(\omega)
=K\omega^2 \int dr\ {\bf a}^2\,.
\eqno(5.29)$$
We showed in section 3
that if the vector potential is written in the form (5.27),
then the potential $\varphi$ is zero when the conductivity is 
isotropic and local. Inspection of (5.20) shows that the boundary 
contribution to the correlation function vanishes when $\varphi $ 
vanishes. In more general cases, comparison of (5.28) 
and (5.29) shows that if the electric field is independent
of frequency, the absorption coefficient is reduced,
relative to its classical value, at frequencies below the
Thouless frequnency $\omega_{\rm c}$.
\gap\gap\gap
\vfill
\eject
%
%
%
%
\noindent {\bf 6. Self-consistent electric field: diffusive case}
\gap
\noindent{\sl 6.1 Non-local conductance}
\gap
We can deduce the non-local conductivity from the results of the 
previous section in two ways. We could use (4.6) as the definition
of the conductivity, and evaluate it by setting ${\bf A}({\bf r})=
\delta ({\bf r}-{\bf R}){\bf e}_i$ in (5.5), so that
$$\sigma_{ij}({\bf R},{\bf R}';t)=
{e^2\over{(2\pi \hbar)^d\delta t^2}}
\langle \delta ({\bf R}-{\bf r})\delta r_i 
\delta ({\bf R}'-{\bf r}')\delta r_j\rangle
\ .\eqno(6.1)$$
Alternatively, if we write the absorption coefficient in terms
of the non-local conductivity in the form
$$\alpha (\omega)={1\over{2E_0^2}}\int d{\bf r}\int d{\bf r}'
\int d\tau \exp[{\rm i}\omega \tau]\,
\sigma_{ij}({\bf r},{\bf r}',\omega)\,E_i({\bf r})\,E_j({\bf r}')
\eqno(6.2)$$
the kernel $\sigma_{ij}({\bf r},{\bf r}';t)$ is deduced
from (5.18). By either route we find
$$\sigma({\bf r},{\bf r}';t)=\Sigma_0
\Big[\delta_{ij}\delta({\bf r}-{\bf r}')
\delta (t)-
 D\partial^2_{r_ir_j'}{\cal P}({\bf r},{\bf r}';t)
\Big]\eqno(6.3)$$
where $\Sigma_0$ is the bulk conductivity, and 
${\cal P}({\bf r},{\bf r}',t)$ is the propagator, given by (5.19).

This form for the non-local conductivity was originally
given in ref. [13] (and in the DC limit in ref. [12]).
The argument in these earlier papers involves the diagrammatic 
analysis of disorder averaged perturbation theory, and in the case
of [13] it appeals to a supersymmetric formalism. We
believe that our derivation is more direct, and also more
compelling. Our derivation considers the effect of the 
surface explicitly, whereas it is not clear from the diagrammatic
analysis that there are not additional contributions which
arise from integrating fields over the surface of the sample. 
Our derivation also deals explicitly with the fact that the
trajectories are discontinuous, and we show explicitly how
the correct evaluation is related to the Stratonovitch
definition of the integral over the trajectory.
\gap
\noindent{\sl 6.2 Self-consistent solution}
\gap
The self-consistent electric field has to be chosen
to satisfy (3.5). 
First we remark that in the case of diffusive electron motion, 
the term containing $\epsilon_0$ is negligible, and can be dropped.
Estimating the magnitude of $\hat \sigma$ by the bulk conductance
$\Sigma_0=ne^2D$, and noting that the bulk plasma frequency scale
is $\omega^2_{\rm p}\sim Ne^2/m_{\rm e}\epsilon_0$, we see that this term 
is negligible provided $\omega\gg \omega_{\rm s}$, where
$\omega_{\rm s}$ is the elastic scattering rate.
This is consistent with the assumptions that the electron
motion is diffusive.

We therefore wish to determine an electric field for which 
$\nabla.(\hat \sigma {\bf E})=0$; more explicitly we require
$$\Sigma_0\,\partial_{r_i}
\int dt\ \exp({\rm i}\omega t)
\int d{\bf r'}\ \bigl[E_i({\bf r}')\delta({\bf r}-{\bf r}')\delta (t)-
D\partial^2_{r_ir'_j}{\cal P}({\bf r},{\bf r}';t)\ E_j({\bf r}')
\bigr]=0\ .\eqno(6.4)$$
Substituting for the propagator using (5.19), we find
$$0=\Sigma_0\biggl[\partial_{r_i} E_i
-\sum_{\alpha}{D^2k_\alpha^2\over{D^2k_\alpha^4+\omega^2}}
\partial_{r_i}\int d{\bf r}'
\partial_{r_i}\chi_\alpha ({\bf r})
\partial_{r'_j}\chi_\alpha ({\bf r}')E_j({\bf r}')\biggr]$$
$$=\Sigma_0\,\partial_{r_i}E_i+\Sigma_0\sum_\alpha
{D^2k_\alpha^4\over{D^2k_\alpha^4+\omega^2}}\,
\chi_\alpha({\bf r})\int \!d{\bf r}'\, \partial_{r_j}
\chi_\alpha({\bf r}')\,E_j({\bf r}')$$
$$=\Sigma_0\,\partial_{r_i}E_i-\Sigma_0\sum_\alpha\
{D^2k_\alpha^4\over{D^2k_\alpha^4+\omega^2}}\,
\chi_\alpha({\bf r})\Big [
\int \!d{\bf r}' \,\chi_\alpha ({\bf r}')
\partial_{r'_j}E_j({\bf r}')$$
$$ 
\;\;\;\;\;\;\;\;\;\;\;\;\;
\;\;\;\;\;\;\;\;\;\;\;\;\;
\;\;\;\;\;\;\;\;\;\;\;\;\;
\;\;\;\;\;\;\;\;\;\;\;\;\;
-\int \!ds_i\, 
\chi_\alpha({\bf s})\,
E_i({\bf s}) \Big]
\ .\eqno(6.5)$$
This equation has a solution where $\nabla.\,{\bf E}=0$
everywhere, with ${\bf E}$ tangential to the boundary.
The classical solution for an isotropic local conductance
satisfies these conditions. We conclude that, at least in the
case where there is no static biasing magnetic field
applied, the electric field configuration is independent 
of frequency.
\gap
\noindent{\sl 6.3 Calculation of the absorption coefficient}
\gap
We have shown that the electric field distribution is 
independent of frequency: we can therefore use the solution
of the form (3.7), (3.8), with $\phi=0$, and with ${\bold \psi}$
tangential to the boundary. Referring to (5.21) we observe
that, after integrating by parts, the integrals are seen to 
vanish because the field ${\bf A}$ is divergenceless and
is tangential at the boundary. It follows that the summation 
in (5.21) vanishes, and that the absorption coefficient
is given by the classical expression (5.1), at all relevant 
frequencies. 

We conclude by describing a useful approach to calculating the
field $\psi $, satisfying Poisson's equation (3.9).
The solution can be obtained from a Green's function 
$G({\bold r},{\bold r}')$ satisfying
$\nabla^2 G=-\delta ({\bold r}-{\bold r}')$. A suitable Green's
function is
$$G({\bold r},{\bold r}')=\sum_n
{\xi_n({\bold r})\xi_n({\bold r}')\over {k_n^2}}\eqno(6.6)$$
where the $\xi_n({\bold r})$ and $k_n^2$ are eigenfunctions and
eigenvalues of the Helmholtz equation, $[\nabla^2+k_n^2]\xi_n=0$,
solved with the Dirichlet boundary condition $\xi_n({\bf r}) =0$.
The field $\psi({\bf r})$ is then obtained by applying this Green's 
function to the source term ${\rm i}\omega B_0$ appearing in
(3.9). In two dimensions, 
the absorption coefficient can then be written in terms of 
the $\xi_n$ as follows
$$\alpha (\omega )
={\Sigma_0\over{2E_0^2}}\int d{\bold r}\ \big|\nabla \psi\big|^2
=-{\Sigma_0\over{2E_0^2}}\int d{\bold r}\ \psi \nabla^2\psi^\ast
={\rm i}\omega{\Sigma_0 B_0\over{2E_0)^2}}
\int d{\bold r}\ \psi
\ .\eqno(6.7)$$
Using the Green's function (6.6) to obtain $\psi $, we have
$$\alpha (\omega )={\Sigma_0\omega^2\over{2c^2}}
\sum_n {1\over {k_n^2}}
\bigg\vert \int d{\bold r}\ \xi_n({\bold r})\bigg\vert^2
\ .\eqno(6.8)$$
This is a very general expression for the classical magnetic
dipole absorption coefficient in particles with diffusive
electron motion, expressed in terms of solutions of the
two dimensional Helmholtz equation. 
As it stands, equation (6.8)
is valid for two-dimensional particles. 

We note that the eigenfunctions $\xi_n({\bf r})$ obey Dirichlet
boundary conditions and that
the absorption coefficient for a given geometry can 
be significantly reduced by applying cuts orthogonal to the boundary:
the main contribution to the sum in (6.8) comes from
the ground state and the low-lying states and
the corresponding eigenvalues are increased by applying cuts.
This behaviour is expected since such cuts inhibit the 
flow of eddy currents which causes the absorption.

\gap
\noindent{\sl 6.4 Some examples for simple geometries}
\gap
In this section we summarize our results for specific
geometries, namely discs, squares, and spheres
(the result for the latter is well known [3], and is included
to establish connections with earlier work).
For two dimensional discs of radius $a$, we have
$\psi = i\omega B_0 r^2/4$. Using (3.10) and (5.1) 
we obtain
$$
\alpha(\omega) = {\pi\over 16} {\Sigma_0 \omega^2 a^4\over c^2}
\ .\eqno(6.9)
$$
This result is easily shown to be consistent with (6.8), using
the fact that the $k_n$ are defined by $J_0(k_n a) = 0$ and
that $\sum_n k_n^{-4} = a^4/32$.
For squares of sidelength $a$, 
we obtain
$$
\alpha(\omega) = {32\over \pi^6} {\Sigma_0 \omega^2 a^4\over c^2}
\sum_{m,n > 0}^{\rm odd} {1\over m^2n^2}{1 \over m^2+n^2}
\ .\eqno(6.10)
$$
The sum can be evaluated numerically and gives $0.528\ldots\ $.
Finally, for spheres of radius $a$
(6.8) is to be modified
as follows
$$
\alpha(\omega) = {\Sigma_0 \omega^2\over 2 c^2}\,
                  \int_{-a}^a\! dz \, 
\sum_n {1\over {k_n^2}}
\bigg\vert \int d{\bold r}\ \xi_n({\bold r})\bigg\vert^2
\eqno(6.11)$$
where the eigenvalues are defined by $J_0(k_n r_\bot)=0$ and
$r_\bot^2 = a^2-z^2$. We obtain
$$
\alpha(\omega) = {\pi\over 15} {\Sigma_0 \omega^2 a^5\over c^2}
\ .\eqno(6.11)
$$
Equation (6.11) reproduces the well-known absorption coefficient
for metallic spheres [3]. In [3], this result
is compared to the absorption coefficient for electric
dipole absorption, which has a different size dependence,
$\sim a^3$. In two dimensions, on the other hand, the
size dependence is the same for magnetic and
electric dipole absorption. The latter coefficient was
calculated in [9],
$$
\alpha_{\rm el}(\omega)
={34\over 9\pi}{\epsilon_0^2 a^4 \omega^2\over\Sigma_0}
\ . \eqno(6.12)
$$
\vfill
\eject
%
%
%
%
\noindent{\bf 7. Ballistic electron motion}
\gap
\noindent{\sl 7.1 General remarks}
\gap
In the case where the electron motion is ballistic, the
electric field must be determined by the non-local
conductivity, and equations (3.5) and (3.6) must be
solved to determine the electric field. Fortunately,
in both the high and low frequency limits, there
are considerations which simplify the discussion.

In the low frequency limit, $\omega \ll \omega_{\rm c}$, we showed
in section 3 
that only $\nabla \wedge {\bold E}$ is relevant, and that we can
use any electric field for which the circulation is spatially
uniform. It follows from (2.3)-(2.5) that the absorption
coefficient is proportional to $\omega^2$ at low frequencies.

In the high frequency limit, the conductivity tensor (4.10) acts
over a range $R\sim v_{\rm F}/\omega$, and the conductivity 
becomes effectively local, with value 
$\Sigma (\omega)={\rm i}Ne^2/m_e\omega$.
The non-local self consistency condition for the electric field
then reduces to the same requirements as for the diffusive case:
the electric field is tangential to the boundary, and is derived
from a field $\psi $ which satisfies (3.9). 

>From (2.5), it is clear that the high 
frequency behaviour is determined
by discontinuities in derivatives of the correlation function.
The correlation function of the smooth perturbation 
$f(t)=\dot {\bf r}.{\bf E}$ for motion in a billiard has discontinuous 
derivatives due to the change of direction when the particle
collides with the boundary.
In the neighbourhood of a collision with the boundary at $t=0$,
the perturbation takes the form
$$f(t)=
 (\dot {\bf r}'+\dot {\bf r}'').{\bf E}[(\dot{\bf r}'+\dot {\bf r}'')t]
         \Theta(-t)
+(\dot {\bf r}'-\dot {\bf r}'').{\bf E}[(\dot{\bf r}'-\dot {\bf r}'')t]
         \Theta(t)
\eqno(7.1)$$
where $\dot {\bf r}'$ and $\dot {\bf r}''$ are respectively tangential
and normal components of the velocity at the instant before the collision,
and $\Theta (t)$ is the step function. Taylor expanding ${\bf E}({\bf r})$ we
find (with repeated indices summed over):
$$f(t)\sim\dot r_i'\, E_i+\dot r_i''\,E_i\,\{\Theta(-t)-\Theta(t)\}
+(\dot r_i'\dot r_j'+\dot r_i'' \dot r_j'')\,\partial_{r_j}E_i\,t$$
$$+(\dot r_i'\dot r_j''+\dot r_i'' \dot r_j')\,\partial_{r_j}E_i\,
t\,\{\Theta(-t)-\Theta(t)\}\eqno(7.2)$$
For a general electric field, $f(t)$ has discontinuities in $t$
of magnitude $2\,\dot r_i''\,E_i$ on collision with the boundary,
but for electric fields tangential to the boundary,
the discontinuities are in the first derivative, and are of
magnitude $2\,(\dot r_i'\dot r_j''+\dot r_i''\dot r_j')\,\partial_{r_j}E_i$.

If $f(t)$ has discontinuities in its $n^{\rm th}$ derivative,
the Fourier transform of its correlation function decays
as $\omega^{-2(n+1)}$ as $\omega \to \infty$. The absorption
coefficient is obtained from this Fourier transform by
multiplying by a factor which contains $\omega^2$. In the
case of a general field we therefore expect the absorption
coefficient to approach a constant for $\omega \gg \omega_{\rm c}$,
whereas for a tangential field we expect that 
$\alpha(\omega)\sim \omega^{-2}$ for $\omega \gg \omega_{\rm c}$.
We know that the field ${\bf E}({\bf r},\omega)$ approaches 
the tangential form as $\omega \to \infty$, but we have
no information about how rapidly this limit is appraoched.
We can only say that the absorption coefficient must
decrease for $\omega \gg \omega_{\rm c}$, and that it
is unlikely to decrease faser than $\omega^{-2}$.
\gap
\noindent{\sl 7.2 An example: the square billiard}
\gap
It is instructive to discuss an example: we consider the 
absorption coefficient for a square billiard with ballistic
electron motion, with two different, frequency independent, 
choices for the electric field ${\bf E}({\bf r})$, both satisfying
(3.7), (3.9). First we calculate the absorption coefficient assuming 
that the field is tangential to the boundary, and 
then consider the case where it is circularly 
symmetric (which corresponds to taking the 
angular momentum operator $\hat L_z$ as the perturbation.
The results will illustrate the application of (2.5),
and will verify two of the conclusions from the arguments
above: we find that the absorption coefficients agree
in the low frequency limit, and that at high frequencies 
the absorption scales as $\omega^{-2}$ for the tangential
field, but as $\omega^0$ for the radially symmetric field.
 
In terms of the perturbation 
$$\Delta H({\bf r},{\bf p}) = {e\over m_{\rm e}} {\bf p}.{\bf A}({\bf r})
\eqno(7.3)
$$
the absorption coefficient is given by
$$
\alpha(\omega) = {\omega^2 \over 2} g(E_{\rm F}) 
{\roman{Re}}\int_0^\infty \!dt  \;
{\roman{e}}^{i\omega t} \;
\Big\langle \Delta H({\bf r}_t,{\bf p}_t) \,
            \Delta H({\bf r},{\bf p})\Big\rangle\,.
\eqno(7.4)
$$
The perturbation (7.3)  is determined by the choice of
the vector potential ${\bf A}({\bf r})$. We will first assume
that ${\bf A}({\bf r})$ is tangential to the boundary of the
particle. Accordingly we take $\phi =0$ in equation (3.7).
The field $\psi(x,y)$ in equation (3.8) is determined from 
equation (3.9) which
is most conveniently solved using the Green's function (6.6).
For a square of side $a$ we obtain for the perturbation
$$ 
\Delta H({\bf r},{\bf p}) 
= {e\over m_{\rm e}} B_0 a^2 \Big({2\over \pi}\Big)^4 \Bigg[
p_x \sum_{mn}^{\roman{odd}} {n\pi\over a}
{\sin\big({m\pi x/a}\big) \cos\big({n\pi y/ a}\big) \over
mn(m^2+n^2)}
$$
$$\hskip 2.75cm +p_y \sum_{mn}^{\roman{odd}} {m\pi\over a}
{\cos\big({m\pi x /a}\big) \sin\big({n\pi y/ a}\big) \over
mn(m^2+n^2)}\Bigg]
\ .\eqno(7.5)
$$
Since motion in the square is integrable, the autocorrelation
function of the perturbation in (7.4) is calculated as an average
over tori
$$
\Big\langle \Delta H({\bf r}_t,{\bf p}_t) \,
            \Delta H({\bf r},{\bf p})\Big\rangle
 = \int \!{d^2\theta \over (2\pi)^2}\;d\mu({\bf I})\;
        \Delta H({\bf I},\boldtheta)\;
        \Delta H({\bf I},\boldtheta+\boldomega({\bf I}) t)
\eqno(7.6)$$
where $d\mu({\bf I}) = g(E_{\rm F})^{-1} \delta[E_{\rm F}-H({\bf I})]$
averages over the tori. ${\bf I}$ and $\boldtheta$ are the action
and angle variables  characterizing the motion in the square,
$\boldomega({\bf I})$ are the respective frequencies.
Equation (7.6) is easily evaluated [19]. The result
is of the form
$$
\alpha(\omega) = {8\over\pi^8}
                 {m_{\rm e} e^2  \omega^2 a^5 v_{\rm F}\over c^2 \hbar^2}\,
                 f(\omega/\omega_{\rm c})
\eqno(7.7)$$
where $f(z)$ is an energy-independent scaling function
and $\omega_{\rm c} = v_{\rm F}/a$. For large frequencies, 
$f(z) \sim z^{-4}$ and hence $\alpha(\omega) \sim \omega^{-2}$.
For small frequencies, on the other hand, one obtains
$$
\alpha(\omega) = {8\over\pi^8}
                 {m_{\rm e} e^2  \omega^2 a^5 v_{\rm F}\over c^2 \hbar^2}\,
\sum_{m,n > 0}^{\rm odd} {1\over m^2n^2}{1 \over (m^2+n^2)^{3/2}}\,.
\eqno(7.8)
$$
As remarked
in section 7.1, the absorption coefficient is proportional 
to $\omega^2$ for small frequencies.

In order to verify explicitely that the low-frequency absorption
does not depend on the boundary conditions for the
electric field, as discussed in sections 3.3 and 7.1, the above
calculation can be repeated
using a vector potential in the symmetric gauge
$${\bf A} = {B_0\over 2} (-y,x,0)
\ .\eqno(7.9)$$
We note that this choice of the vector potential
does note satisfy tangential boundary conditions.
The corresponding perturbation is
$$\Delta H({\bf r},{\bf p}) = {e\over m_{\rm e}} {\bf p}.{\bf A}
                   = {e B_0\over 2 m_{\rm e}} L_z\ .\eqno(7.10)$$
For small frequencies we find again the result (7.8),
thus verifying explicitly that the boundary conditions
do not influence the low-frequency absorption.
For high frequencies, on the other hand,
we find $\alpha(\omega) \sim \omega^0$, as predicted 
in the previous section.
\gap\gap\gap
\vfill
\eject
%
%
%
%
\noindent{\bf 8. Acknowledgements}
\gap
We are grateful to the organisers of the program on
\lq Quantum Chaos and Disordered Systems' for inviting us to
the Newton Institute. The work of BM was supported by the Max Planck
Institute for Physics of Complex Systems, and that of MW was supported by a
research grants from the EPSRC, reference GR/H94337
and GR/L02302. PNW was working at Strathclyde University
when this project was started, where his work 
was supported by an EPSRC grant;
he is currently supported by a EU fellowship.
\gap\gap\gap
\vfill
\eject
%
%
%
%
\vfill
\eject
{\indentoff \bf References}
\gap
\indentoff
\gap
\ref {1}{G. Mie}{Ann. Physik.}{25}{377}{1908}
\gap
[2] N. W. Ashcroft and N. D. Mermin, {\sl Solid State Physics},
Philadelphia: Saunders College, (1976).
\gap\gap
[3] E. M. Lifshitz and L. P. Pitaevskii, {\sl Statistical Physics, Part 2},
(Volume 9 of Landau and Lifshitz Course of Theoretical Physics),
Oxford: Pergamon, (1980).
\gap
\ref{4}{D. B. Tanner, A. J. Sievers and R. A. Buhrman}
{Phys. Rev.}{B11}{1330-41}{1975}
\gap
\ref{5}{D. Stroud and F. P. Pan}{Phys. Rev.}{B17}{1602-10}{1978}
\gap
[6] G. L. Carr, S. Perkowitz, and D. B. Tanner, in {\sl Infrared and
Millimeter Waves}, {\bf 13}, 169, ed. K. J. Button, Academic Press, (1985).
\gap\gap
\ref {7}{E. J. Austin and M. Wilkinson}{J. Phys.: Condens. Matter}
         {5}{8461-8484}{1993}
\gap
\ref {8}{M. Wilkinson and E. J. Austin}{J. Phys.: Condens. Matter}
         {6}{4153-4166}{1994}
\gap
\ref {9}{B. Mehlig and M. Wilkinson}{J. Phys.: Condens. Matter}
         {9}{3277-90}{1997}
\gap
[10] L. P. Gorkov and G. M. Eliashberg, {\sl Zh. Eksp. Teor. Fiz.}, 
{\bf 48}, 1407, (1965) ({\sl transl.} {\sl Sov. Phys. JETP}, {\bf 21}, 
940, (1965)).
\gap\gap
\ref{11}{N. Argaman}{Phys. Rev.}{B47}{4440}{1993}
\gap
\ref{12}{C. L. Kane, R. A. Serota and P. A. Lee}{Phys. Rev.}{B37}{6701-10}
{1988}
\gap
\ref{13}{R. A. Serota, J. Yu and Y. H. Kim}{Phys. Rev.}{B42}{9724-7}{1990}
\gap
\ref{14}{M. Wilkinson}{J. Phys.}{A20}{2415}{1987}
\gap
\ref {15}{D. J. Thouless}{Phys. Rep.}{13}{93}{1974}
\gap
\ref{16}{A. Kamenev and Y. Gefen}{Int. J. Mod. Phys.}{B9}{751-802}{1995}
\gap
\ref{17}{B. Reulet, M. Ramin, H. Bouchiat and D. Mailly}
{Phys. Rev. Lett.}{75}{124-27}{1996}
\gap
[18] N. G. van Kampen, {\sl Stochastic processes in physics and chemistry}, 
Amsterdam: North Holland, (1981).
\gap\gap
\ref{19}{B. Mehlig}{Phys. Rev.}{B55}{R10193}{1997}
\vfill
\eject
%
%
%
\noindent{\bf Figure captions}
\gap\gap
Figure 1: Illustrating the images used in discussing 
expression (4.11) for a square billiard. The sum over
all paths from ${\bf r}$ to ${\bf r}'$ can be represented
as a sum over straight lines $k$ of length $L_k$ connecting ${\bf r}$
with the image points of ${\bf r}'$.
\gap
Figure 2: Illustrating the vectors and coordinate system used
in the discussion of the construction of the short time propagator.
\vfill
\eject
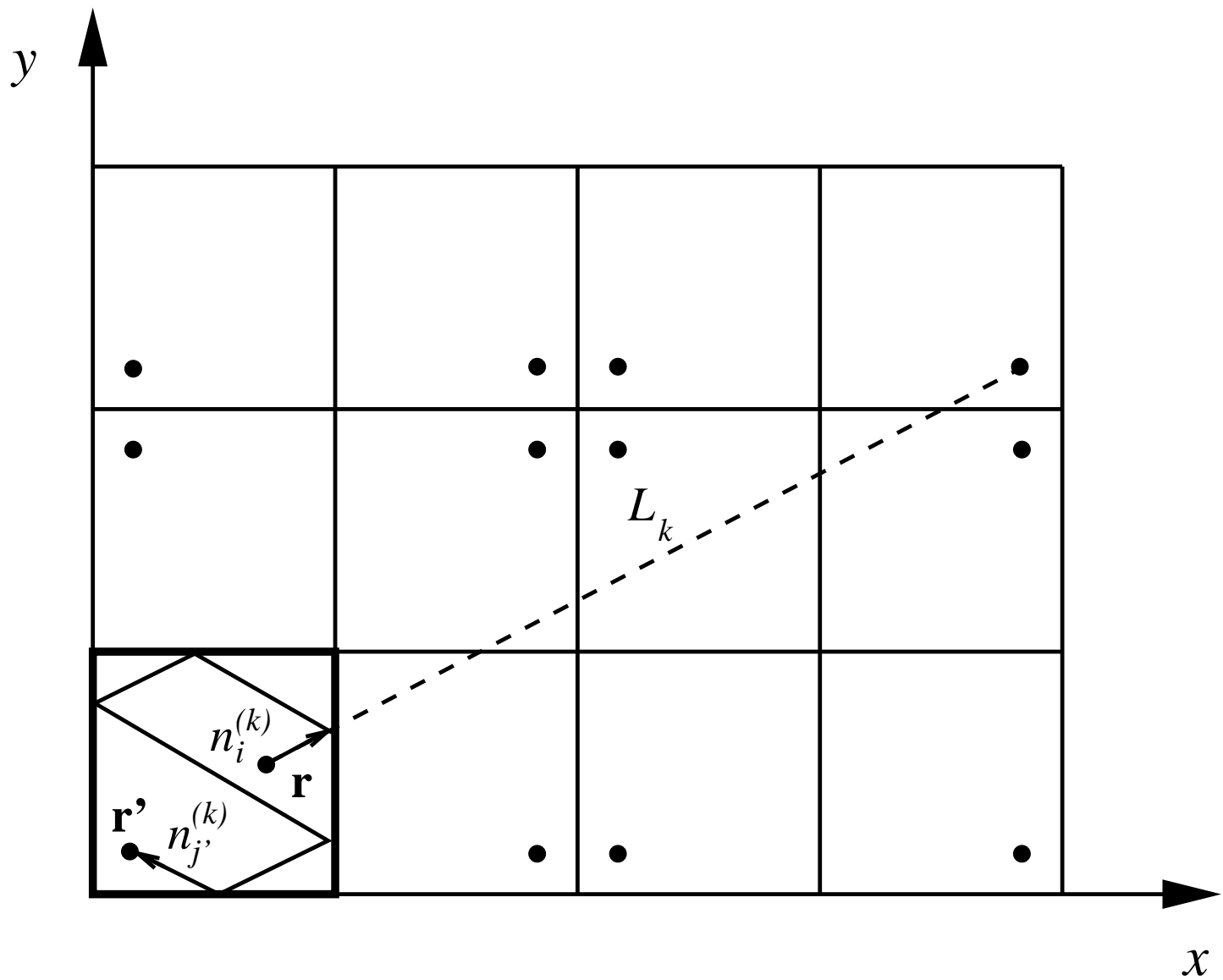
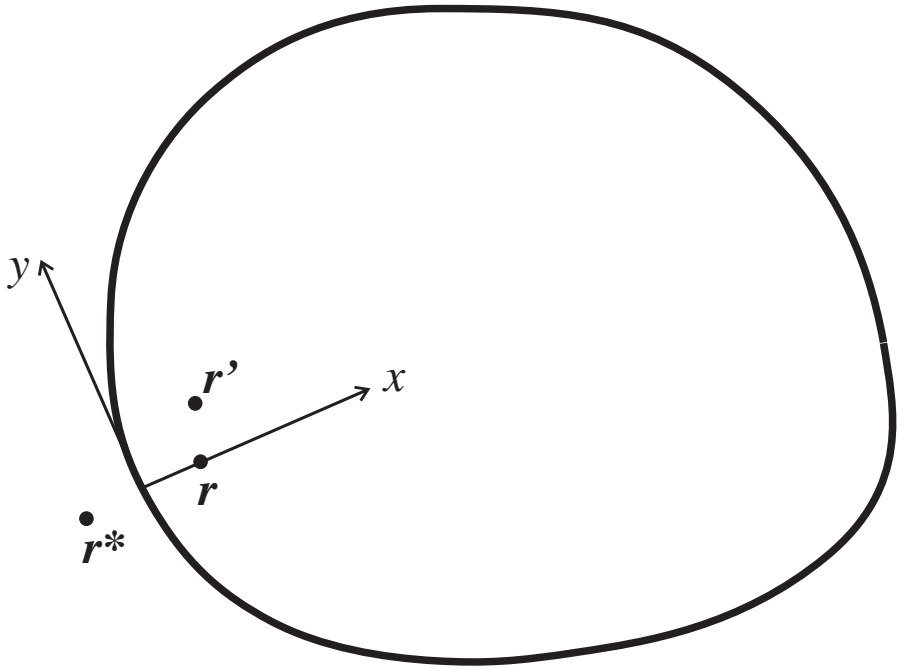
\end